\documentclass[structabstract]{aa}
\usepackage{natbib}
 \usepackage{hyperref}
  \hypersetup{colorlinks=true,citecolor=blue}
\usepackage{graphicx}
\usepackage{draftcopy}
\newcommand{\be}{\begin{equation}}
\newcommand{\ee}{\end{equation}}
\newcommand{\ba}{\begin{align}}
\newcommand{\ea}{\end{align}}
\newcommand{\bea}{\begin{eqnarray}}
\newcommand{\eea}{\end{eqnarray}}
\newcommand{\rd}{\rm{d}}

\title{3D Spherical Analysis of Baryon Acoustic Oscillations}
\author{A. Rassat $^{1,2}$ \and A. Refregier $^3$}
\institute{$^1$ Laboratoire d'Astrophysique, Ecole Polytechnique F\'ed\'erale de Lausanne (EPFL), Observatoire de Sauverny, CH-1290, Versoix, Switzerland.\\
$^2$ Laboratoire AIM, UMR CEA-CNRS-Paris 7, Irfu, SAp/SEDI, Service d'Astrophysique, CEA Saclay, F-91191 GIF-SUR-YVETTE CEDEX, France. \\ 
$^3$ Institute for Astronomy, ETH Z\"urich, Wolfgang Pauli Strasse 27, CH-8093 Z\"{u}rich,  Switzerland.\\}

\abstract{Baryon Acoustic Oscillations (BAOs) are oscillatory features in the galaxy power spectrum which are used as a standard rod to measure the cosmological expansion. These have been studied in Cartesian space (Fourier or real space) or in spherical harmonic space in thin shells.}{Future wide-field surveys will cover both wide and deep regions of the sky and thus require a simultaneous treatment of the spherical sky and of an extended radial coverage.  The Spherical Fourier-Bessel (SFB) decomposition is a natural basis for the analysis of fields in this geometry and facilitates the combination of BAO surveys with other cosmological probes readily described in this basis. In this paper, we present a new way to analyse BAOs by studying the BAO wiggles from the SFB power spectrum.}
{In SFB space, the power spectrum generally has both a radial ($k$) and tangential ($\ell$) dependence and so do the BAOs. In the deep survey limit and ignoring evolution, the SFB power spectrum is purely radial and reduces to the Cartesian Fourier power spectrum. In the opposite limit of a thin shell, all the  information is contained in the tangential modes described by the 2D spherical harmonic power spectrum.}{We find that the radialisation of the SFB power spectrum is still a good approximation even when considering an evolving and biased galaxy field with a finite selection function.  This effect can be observed by all-sky surveys with depths comparable to current surveys. We also find that the BAOs radialise more rapidly than the full SFB power spectrum. }{Our results suggest the first peak of the BAOs in SFB space becomes radial out to $\ell \sim 10$ for all-sky surveys with the same depth as SDSS or 2dF, and out to $\ell \sim 70$ for an all-sky stage IV survey. Subsequent BAO peaks will also become radial, but for shallow surveys these may be in the non-linear regime.  For modes that have become radial, measurements at different $\ell$'s are useful in practice to reduce measurement errors.}

\begin{document}
\titlerunning{3D Spherical Analysis of BAOs}
\maketitle

\section{Introduction}

The study of Large Scale Structure (LSS) with galaxy surveys is a promising tool to study the dark universe \citep{WGFC,DETF}. Baryon Acoustic Oscillations (BAOs) are a special feature in the galaxy power spectrum present on scales $100h^{-1}$Mpc, which are due to oscillations in the coupled baryon-photon fluid before recombination \citep{Sunyaev:1970Z,Peebles:1970Y,Eisenstein:2005,Seo:2003Eis, Seo:2007Eis}. The BAOs are considered a powerful cosmological tool as the BAO scale acts as a standard ruler with which to probe cosmic expansion both in the radial and tangential directions. 

BAOs were first detected by \cite{Eisenstein:2005} in SDSS data \citep{SDSS} and later with 2dF galaxies \citep{2dFdata:2003,Cole:2005} and finally with both surveys \citep{Percival:2007}, though others suggest current data cannot currently probe the BAO scale sufficiently \citep{Cabre:Gaz}.

Until now, BAOs have been studied in Cartesian space, either in Fourier space \citep{Seo:2003Eis,Seo:2007Eis}, or in real space \citep{Eisenstein:2005,baoconfig1,baoconfig2}, and in 2D spherical harmonic space on thin spherical shells  \citep{Dolney:2006}. These descriptions use different information and therefore have different constraining power for cosmological parameters \citep{Rassat:2008bao}. 

Future wide-field BAO surveys will, however, cover both large and deep areas of the sky, and thus require a simultaneous treatment of the spherical sky geometry and of extended radial coverage. The Spherical Fourier-Bessel (SFB) decomposition is a natural basis for the analysis of fields in this geometry.  The SFB analysis is powerful as it uses a coordinate system in which the radial selection function and physical effects are naturally described. Moreover, this description facilitates the combination of BAO surveys with other cosmological probes which are readily described in the SFB decomposition such as the smooth power spectrum and redshift space distortions \citep{Heavens:1995,Fisher:1995,Percival:2004,Erdogdu:2005wi,Erdogdu:2006dv}, weak lensing \citep{Heavens:2003,CHK:2005,Kitching:2008}, and the Integrated Sachs-Wolfe effect \cite[ISW,][]{ISW3D}. Studying BAOs from wide-field surveys with an SFB expansion,  is therefore natural both for the geometry considered, and for unifying the treatment with the other probes. 

In \S\ref{sec:theory} we first review the general decomposition of a 3D field in SFB space and introduce the concept of \emph{radialisation} when the field is statistically isotropic and homogeneous. In \S\ref{sec:3dBAO:1}, we introduce a new way to consider the BAOs, by studying the baryon wiggles in SFB space, in a similar way as in Fourier space, while in \S\ref{sec:radialisation} we discuss the radialisation of BAOs in SFB space in the context of existing and future surveys. In \S\ref {sec:conclusion} we present our conclusions. 

\section{3D Spherical Fourier-Bessel (SFB) Expansion}\label{sec:theory}

\subsection{Expansion of an homogeneous and isotropic field}\label{sec:theory:nonevol}
Let us consider a field $f({\bf r})$ at time $t$, where ${\bf r}=(r, \theta, \varphi)$ in spherical polar coordinates. In practice, this field may represent the galaxy or mass density (or overdensity) in the universe. In a flat geometry, the field
can be decomposed in the 3D SFB basis set, which is complete and orthonormal, as
\begin{equation} f({\bf r})=\sqrt{\frac{2}{\pi}}\int {\rd} k \sum_{\ell m} f_{\ell m}(k) kj_\ell(kr) Y_{\ell m}(\theta, \varphi),\end{equation}
where $j_{\ell}(x)$ are spherical Bessel functions of the first kind, $Y_{\ell m}(\theta, \varphi)$ are spherical harmonics,
$\ell$ and $m$ are multipole moments and $k$ is the wavenumber. The inverse relation is
\begin{equation} f_{\ell m}(k) = \sqrt{\frac{2}{\pi}}\int {\rd}^3{\bf r}~f({\bf r})kj_\ell(kr)Y^*_{\ell m}(\theta, \varphi), \end{equation}
where we use the same conventions as \cite{Leistedt:2011} and \cite{CHK:2005} \cite[see also][who use a different convention and basis set]{Heavens:1995,Fisher:1995}. 

This decomposition can be viewed as the spherical polar coordinate analogue to the Fourier decomposition in Cartesian
coordinates given by
\begin{eqnarray}
f({\bf x}) & = & \frac{1}{(2\pi)^3}\int {\rd}^3k~\tilde{f}({\bf k}) e^{i{\bf k}\cdot {\bf x}}, \\
\tilde{f} ({\bf k}) & = & \int {\rd}^3x~f({\bf x}) e^{-i{\bf k}\cdot {\bf x}}.
\end{eqnarray}

The 3D SFB power spectrum $C_{\ell}(k)$ of the field $f({\bf r})$ is given by
the the 2-point function of the SFB coefficients $f_{\ell m}(k)$ which
can be written as
\begin{equation} \langle f_{\ell m}(k)f^*_{\ell ' m'}(k')\rangle=C_{\ell}(k)\delta(k-k')\delta_{\ell \ell '}\delta_{m m'},
\label{eq:clk_def}
\end{equation}
when the field is Statistically Isotropic and Homogeneous (hereafter, the SIH condition). Similarly and in the same condition, the Fourier
power spectrum $P(k)$ is implicitly defined by 
\begin{equation} \langle \tilde{f}({\bf k})\tilde{f}^*({\bf k'}) \rangle =(2\pi)^3 P(k) \delta^3({\bf k} - {\bf k'}).\label{eq:pk}\end{equation}

These two power spectra are related by \citep[see for e.g.,][]{Fisher:1995,CHK:2005}:
\begin{equation} C_{\ell}(k) = P(k). \label{eq:clkpk}\end{equation}
Thus, the SFB power spectrum $C_{\ell}(k)$ is independent of the multipole $\ell$, and thus only has a radial ($k$) dependence. 
This remarkable yet recondite fact is only true if the field fulfills the SIH condition. 

In the following subsections, we discuss the impact of the (partial) violation of this condition
in practice and the implication for the measurements of BAO in SFB analysis.

\subsection{Finite Surveys}
In practice, a cosmological field, such as the galaxy density field, will only be partially observed due the finite
survey volume. In this case the observed field $f^{\rm obs}({\bf r})$ can be described by
\begin{equation} f^{\rm obs}({\bf r}) = \phi(r) f({\bf r}),\end{equation} where $\phi(r)$ is the radial selection function
of the survey, and $f({\bf r})$ is assumed to fulfill the SIH condition for now (we will address the
effect of bias and evolution in \S\ref{subsec:evolution}). The observed field $f^{\rm obs}({\bf r})$ is thus no longer
homogeneous because of the radial the selection function. There may also be a tangential selection function to account for regions of missing data, but we assume here that the data is available on the full sky. For convenience, the selection function is normalised as
\begin{equation} \int {\rd}^3 r~\phi(r) = V, \label{eq:phi_norm}\end{equation}
where $V$ is a characteristic volume of the survey chosen such that $\phi \to 1$ as $V \to \infty$, such that 
\be \lim_{V \to \infty} f^{\rm obs}({\bf r}) =f({\bf r}).\ee

In this case, the homogeneity condition (SIH) is no longer valid in the radial direction and the observed 2-point function
can be written as 
\begin{equation} \langle f^{\rm obs}_{\ell m}(k)f^{\rm obs*}_{\ell ' m'}(k')\rangle=C^{\rm obs}_{\ell}(k,k')\delta_{\ell \ell '}\delta_{m m'}
\end{equation}
where the observed SFB power spectrum $C^{\rm obs}_{\ell}(k,k')$ depends this time on $\ell$, $k$ and $k'$, as opposed to only on $\ell$ and $k$ as in Equation \ref{eq:clk_def}.
It can be shown that it can be expressed as
\begin{equation}
C^{\rm obs}_{\ell}(k,k') = \left(\frac{2}{\pi}\right)^2 \int {\rd} k'' ~k''^2 P(k'') W_\ell(k,k'') W_\ell(k',k'')
,\label{eq:clk_galfield}\end{equation}
where $P(k)$ is defined in Equation \ref{eq:pk} and the window function $W_\ell(k,k')$ is defined as
\begin{equation} W_\ell(k,k') = \int {\rd} r~r^2 \phi(r) kj_\ell(kr)j_\ell(k'r). \label{eq:window}\end{equation}
In practice, $C^{\rm obs}_{\ell}(k,k')$ tends to fall off rapidly away from the diagonal $k=k'$ and we will often only
compute $\bar{C}_\ell^{\rm obs}(k)$ defined by: \begin{equation}\bar{C}^{\rm obs}_{\ell}(k) \equiv C^{\rm obs}_{\ell}(k,k).\label{eq:clkkclk}\end{equation}

We next evaluate the window function and the observed power spectrum for three special cases for the selection function.

\subsubsection{Gaussian selection function}\label{subsec:gaussian}
As a first example, let us consider a Gaussian selection function defined as
\begin{equation} \phi(r) = e^{-\left(r/r_0\right)^2},\label{eq:phi_gauss}\end{equation}
where $r_0$ is a radius parameter and the normalisation obeys
Equation~\ref{eq:phi_norm} (with $V=\pi^{\frac{2}{3}} r_0^{3}$).
In this case, the window function can be integrated analytically, and Equation \ref{eq:window} becomes: 
\begin{eqnarray} W_\ell(k,k') &=& \frac{\pi r_0^2}{4}\sqrt{\frac{k}{k'}}\exp \left[ -r_0^2 \frac{k^2 + k'^2}{4}\right] I_{\ell+\frac{1}{2}} \left(\frac{r_0^2 kk'}{2}\right),\nonumber\\
&&~ \label{eq:analytic:window}
\end{eqnarray} 
where $I_\nu(x)$ is the modified Bessel function of the first kind. This analytical form considerably facilitates the evaluation of Equation \ref{eq:clk_galfield} (see \S\ref{app:gauss} for a way to numerically calculate the above Equation for large arguments of $I_\nu(x)$).

\subsubsection{Radial Limit}\label{subsec:radial}
In the case where the radial selection function corresponds to full radial coverage (i.e. $\phi(r) = 1, \forall r$), the window function becomes
\begin{equation} W_\ell(k,k') = \frac{\pi}{2k'}\delta(k-k'). \label{eq:w_radial} \end{equation} 
Inserting this expression into Equation \ref{eq:clk_galfield}, we recover Equation \ref{eq:clkpk}, namely
\begin{equation} C_{\ell}^{\rm obs}(k,k')=C_{\ell}(k) \delta(k-k')=P(k) \delta(k-k'),\label{eq:clkpk_obs} \end{equation}
meaning that the 3D spherical spectrum is only dependent on radial coordinate $k$, as discussed above. 

As we show in the appendix \S\ref{app:gauss}, this limit can  also be obtained by taking the limit $r_0 \rightarrow \infty$ for the Gaussian weight function of Equation~\ref{eq:phi_gauss} and is achieved in practice when the 
condition (see Equation~\ref{eq:radial_condition})
\begin{equation} r_0 k  \gg \sqrt{2 \ell (\ell+1)} \label{eq:radial_limit1}\end{equation}
is satisfied. We refer to this as the \emph{radialisation} of the field and discuss this more in \S\ref{sec:3dBAO}. 

\subsubsection{Tangential Limit} \label{subsec:tang}
In the other extreme case, the radial selection function covers only a thin shell of the field at a distance $r_*$
\begin{equation} \phi(r) = r_* \delta(r-r_*),\label{eq:thinshell}\end{equation}
where the normalisation of Equation~\ref{eq:phi_norm} was chosen with $V=4 \pi r_*^3$. Equation 
\ref{eq:clk_galfield} then becomes
\begin{eqnarray} C^{\rm obs}_\ell(k,k')&=& \left(\frac{2}{\pi}\right)^2kk' j_\ell(kr_*)j_\ell(k'r_*)\nonumber\\
&&\times \int {\rd} k'' k''^2 P(k'')\left[r_*^3 j_\ell(k''r_*)\right]^2.\label{eq:clk3D}\end{eqnarray}
 This can be related to the statistics of the 2D projected field 
 \begin{equation} f^{\rm 2D}(\theta, \varphi) =\int {\rd} r r^2 \phi(r) f(r,\theta, \varphi),\end{equation} whose spherical harmonic decomposition is given by \begin{eqnarray} f^{\rm 2D}(\theta, \varphi) &=& \sum_{\ell = 0}^{\infty} \sum_{m=-\ell}^\ell f^{\rm 2D}_{\ell m} Y_{\ell m}(\theta, \varphi),\\
 f^{\rm 2D}_{\ell m} &=&\int {\rd} \Omega f^{\rm 2D}(\theta, \varphi)Y^*_{\ell m}(\theta, \varphi),
 \end{eqnarray}
and whose 2D spherical harmonic power spectrum $C^{\rm 2D}_\ell$ is defined by \begin{equation}  \left< f_{\ell m} f^*_{\ell 'm'}\right>=C^{\rm 2D}_\ell \delta_{\ell \ell'}\delta_{mm'}.\end{equation} 
  
  This power spectrum is related to the 3D Fourier space power spectrum $P(k)$  by:
  \be C^{\rm 2D}_\ell =\frac{2}{\pi}\int {\rd} kk^2P(k)\left[W^{\rm 2D}_\ell(k)\right]^2,\end{equation}
  where the 2D window function $W^{\rm 2D}_\ell(k)$ is given by 
 \be W^{\rm 2D}_\ell(k) = \int {\rd} r r^2 \phi(r) j_\ell(kr). \ee Note that due to our choice of normalisation for $\phi(r)$ in Equation \ref{eq:phi_norm}, $W^{\rm 2D}_\ell(k)$ has units of volume. 
 
 In the case of a thin-shell selection function (Equation \ref{eq:thinshell}), this reduces to
 \be C^{\rm 2D}_\ell=\frac{2}{\pi}\int {\rd} k k^2 P(k) \left[r^3_* j_\ell(kr_*)\right]^2\label{eq:cl2D}.\ee

Thus, the SFB power spectrum (Equation \ref{eq:clk3D}) is simply related to the 2D spherical harmonic power spectrum\footnote{Equation \ref{eq:3d2d} is conceptually different from Equation B3 in \cite{KHM:2010}: the former is an exact solution for a thin shell, while the latter uses the Limber approximation.} by 
\be C^{\rm obs}_\ell(k,k')= \frac{2}{\pi}kk' j_\ell(kr_*)j_\ell(k'r_*) C_\ell^{\rm 2D}.\label{eq:3d2d}\ee
The term $kk'j_\ell(kr_*)j_\ell(k'r_*)$ is a geometric factor which does not depend on the field $f$.  Thus, in this case, all the information in the 3D SFB spectrum $C_\ell(k,k')$ is contained in the 2D power spectrum $C^{\rm 2D}_\ell$, i.e. depends solely on $\ell$.

\begin{figure*}[htbp]
   \centering
   	\includegraphics[width=9cm]{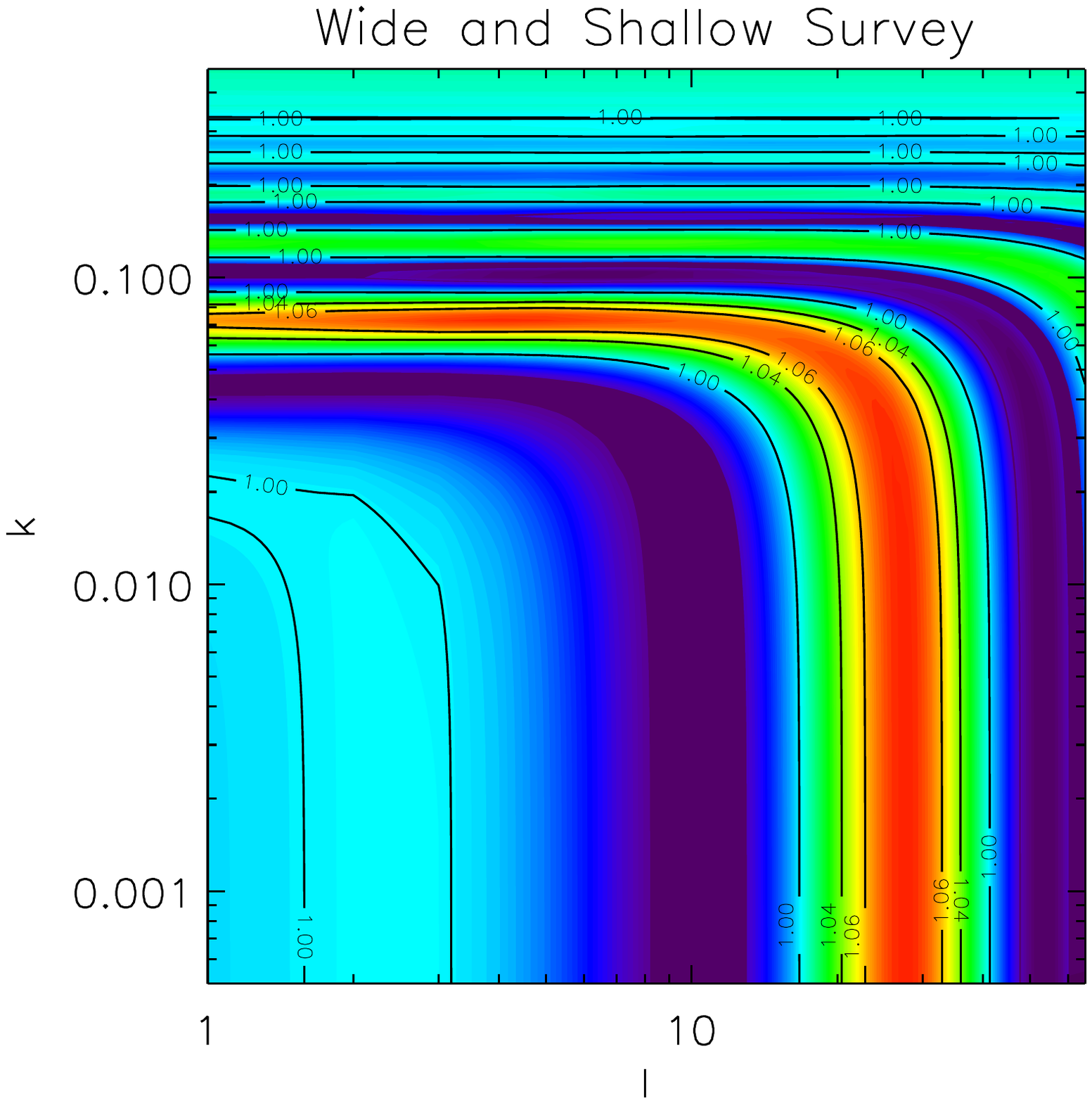}\includegraphics[width=9cm]{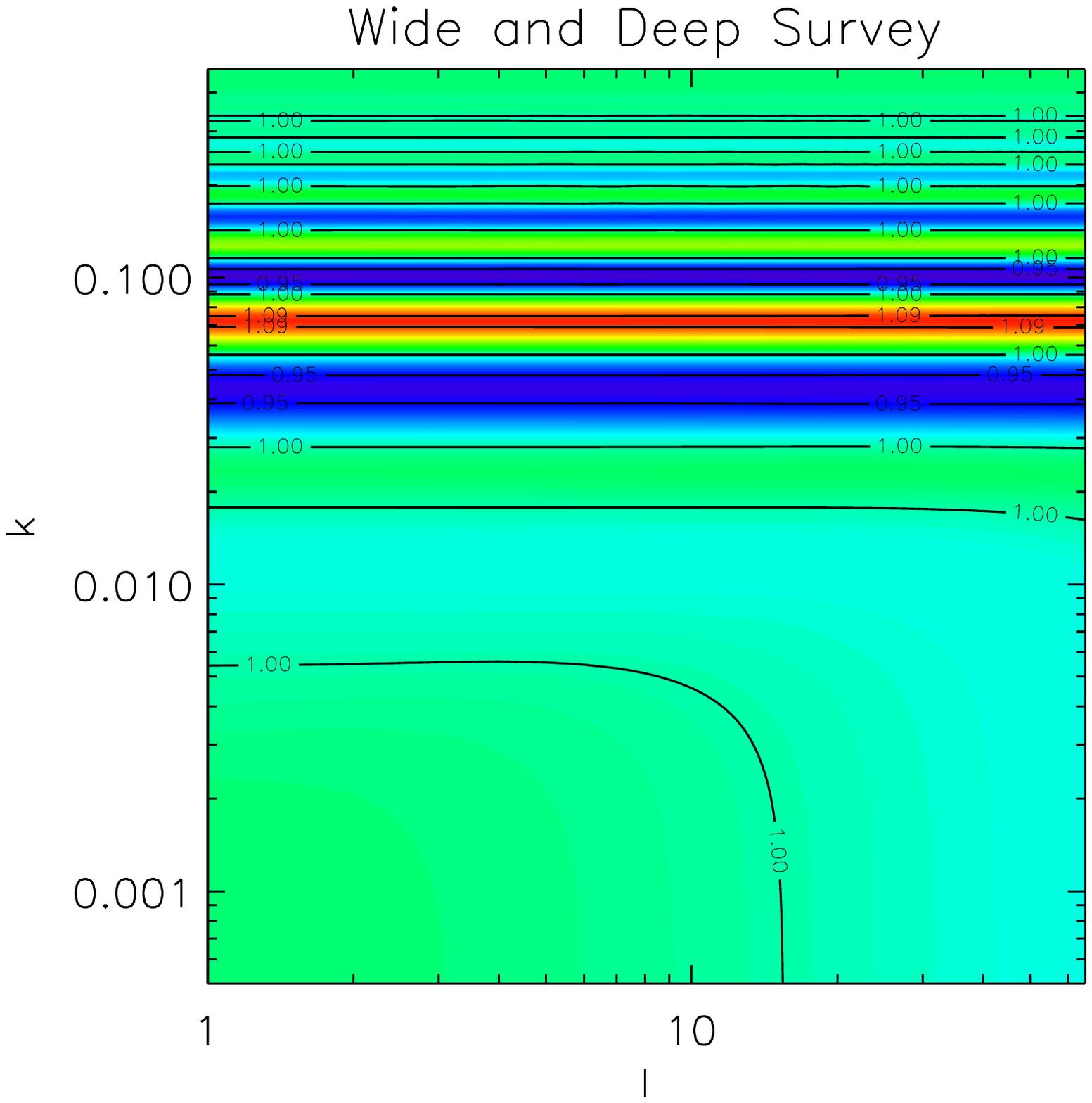}
            \caption{Ratio $R^C_\ell(k)$ of SFB spectrum with and without the physical effects of baryons in $(\ell, k)$ space for a wide and shallow survey (left, $r_0=100h^{-1}{\rm Mpc}$) and a wide and deep survey (right, $r_0=1400 h^{-1}{\rm Mpc}$) using a Gaussian selection function in $r$. The baryonic wiggles are seen both in the radial ($k$) and the tangential ($\ell$) directions, but as $r_0$ is increased, only radial wiggles - with increased amplitude - persist. This \emph{radialisation} of the information is due to mode-cancelling along the line of sight for deep surveys. }
   \label{fig:clk}
\end{figure*}

 \subsection{Evolution and bias}\label{subsec:evolution}

For a realistic galaxy field, the SIH condition will not be met due to the radial evolution of the field
from the growth of structure and redshift dependent bias. 

For a biased and evolving field in the linear regime, the galaxy power spectrum $P^{\rm evol}(k,r)$ will have an explicit distance dependence: \be P^{\rm evol}(k,r) = b^2(r)D^2(r)P(k),\ee 
where the term $b(r)$ is the linear bias, assumed to be scale-independent, $D(r)$ is the growth factor and $P(k)=P(k,r=0)$ is the matter power spectrum at $r=z=0$.  The linear approximation will hold up to a redshift-dependent scale $k_{\rm max}(z)$. In the standard cosmological model, we find $k_{\rm max}(z=0)\simeq 0.12h {\rm Mpc}^{-1}$ (i.e., before the second BAO peak) and $k_{\rm max}(z=2) \simeq 0.25 h{\rm Mpc}^{-1}$. 

For the realistic galaxy field, an approximation for the spherical 3D power spectrum for the galaxy field is still given by Equation \ref{eq:clk_galfield}, but the window function will incorporate the linear growth and bias functions as 

\be W^{\rm evol}_\ell(k',k) = \int {\rd} r ~r^2 \phi(r)k D(r)b(r) j_\ell(kr)j_\ell(k'r). \label{eq:window:realistic}\ee
  
\begin{figure*}[htbp]
   \centering
      \includegraphics[width=11cm]{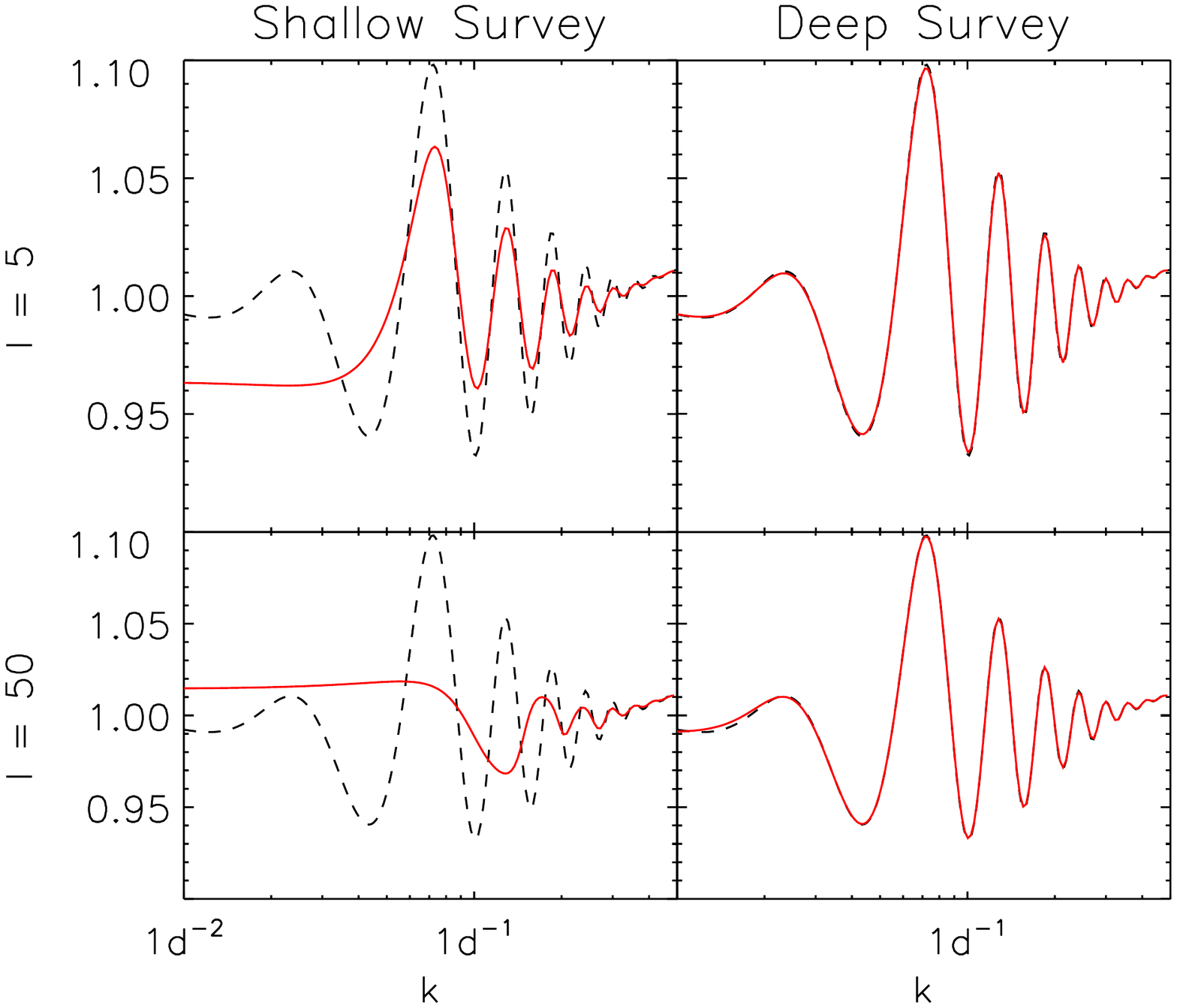}
         \caption{Slices in $R^C_\ell(k)$  (solid, red) compared to $R^P(k)$ (dashed, black) for $\ell=5,50$ (\emph{top, bottom}), for a wide and shallow galaxy survey (left, $r_0=100h^{-1}{\rm Mpc}$, or $z_m \sim0.05$) and a wide and deep galaxy survey (right, $r_0=1400 h^{-1}{\rm Mpc}$ or $z_m \sim 0.8$).}
   \label{fig:slices}
\end{figure*}
\begin{figure}[htbp]
   \centering
   \includegraphics[width=9.5cm]{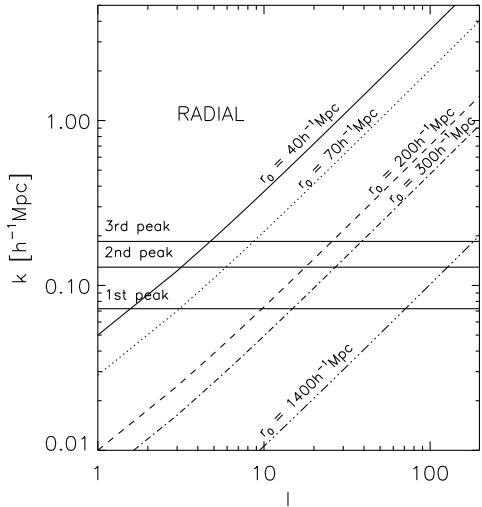}
   \caption{Radialisation limit as defined by Equation \ref{eq:radial_limit1} (diagonal lines) for Gaussian surveys (Equation \ref{eq:phi_gauss}) with $r_0 = 40, 70, 200, 300, 1400 h^{-1}{\rm Mpc}$, i.e. roughly corresponding to the following surveys: IRAS ($z_{\rm mean}\sim0.015$), 2MRS ($z_{\rm mean}\sim0.03$), SDSS ($z_{\rm mean}\sim0.10$), 2dF ($z_{\rm mean}\sim0.15$), and a stage IV survey ($z_{\rm mean}\sim0.80$). Information above the diagonal line is in the radial limit. We overplot the physical scales corresponding to the first three BAO peaks for comparison.}
   \label{fig:radial}
\end{figure}

\section{Application to Baryon Acoustic Oscillations}\label{sec:3dBAO}

\subsection{SFB Analysis of BAOs}\label{sec:3dBAO:1}
The matter power spectrum in Equation \ref{eq:clk_galfield} includes the physical effects of baryons, which create oscillations in Fourier space \citep{Sunyaev:1970Z,Peebles:1970Y,Seo:2003Eis,Eisenstein:2005, Seo:2007Eis}. These Baryon Acoustic Oscillations (BAOs) can be isolated by considering the ratio $R^P(k)$ in Fourier space: 
\be R^P(k) = \frac{P^{\rm b}(k)}{P^{\rm no b}(k)},\label{eq:rp}\ee where $P^{\rm b}(k)$ is the galaxy (or matter) power spectrum including the physical effect of BAOs, and $P^{\rm no b}(k)$ is broad band or `smooth' part of the galaxy (or matter) power spectrum. In linear theory the growth and bias terms cancel out so that Equation \ref{eq:rp} is independent of $z$.
These oscillations can also be probed in 2D spherical harmonic space \citep{Dolney:2006}. We are interested here to see if they can similarly be probed in SFB space as well. 

In SFB space, we consider the ratio $R^C_\ell(k)$ given by: 
 \be R^C_\ell(k) = \frac{\bar{C}^{\rm obs, b}_\ell(k) }{\bar{C}^{\rm obs,no b}_\ell(k)},\label{eq:rcb}\ee where similarly to Fourier space, $\bar{C}^{\rm obs,b}_\ell(k)$ is the diagonal SFB power spectrum (Equation~\ref{eq:clkkclk}) including the physical effects of baryons and $\bar{C}^{\rm obs,no b}_\ell(k)$ is the `smooth' part of the SFB power spectrum. These are calculated by using $P^{\rm b}(k)$ or $P^{\rm nob}(k)$ instead of $P(k)$ in Equation \ref{eq:clk_galfield} and then considering Equation \ref{eq:clkkclk}. The SFB decomposition suggests that the BAOs can manifest themselves in both $k$ and $\ell$ space.

In Figure \ref{fig:clk}, we plot $R^C_\ell(k)$ for two wide-field galaxy surveys with a shallow (left) and deep (right) galaxy Gaussian selection functions with $r_0 = 100 h^{-1}{\rm Mpc}$, and $1400h^{-1} {\rm Mpc}$ respectively (see \S\ref{subsec:gaussian}). We chose a fiducial concordance cosmology with $\Omega_m =0.25 $, $\Omega_{\rm DE} =0.75 $,$\Omega_b =0.045$, $w_0=-0.95$, $w_a =0.0$, $h=0.70$, $\sigma_8=0.80$, $n_s=1$ and no running spectral index. We consider both to be `realistic' surveys, i.e. with evolution due to growth taken into account and a linear galaxy bias (taken as $b(z) = \sqrt{1+z}$), using Equation \ref{eq:window:realistic} for the window function.

In the LHS of Figure \ref{fig:clk} (narrow survey), the BAOs depend on both $k$ and $\ell$ modes, illustrating that the wiggles can be measured simultaneously in the $\ell$ and $k$ directions, which is the first result of this paper. 

However, in the RHS of Figure \ref{fig:clk} (wide survey), the BAOs appear to have only a radial ($k$) dependence, except at very large physical scales ($k<0.01 h{\rm Mpc^{-1}}$), where the dependence is both radial and tangential.  This is a practical illustration of Equation \ref{eq:clkpk_obs} (see also Equation \ref{a5}). This can be understood as tangential modes being attenuated due to mode-cancelling along the line of sight - the attenuation becomes important when the selection function spans a wide redshift range. We refer to this effect as the \emph{radialisation} of the power spectrum, where we have used the baryonic wiggles to probe the radialisation. We discuss the radialisation limit in more detail in \S\ref{sec:radialisation} and in \S\ref{sec:conclusion}.

One of the important results of Figure \ref{fig:clk} is that the radial limit is still reached even when evolution (growth and bias) of the field are considered. This can only be considered numerically since there is no analytic solution to Equation \ref{eq:window:realistic} due to the inclusion of the growth and bias terms. We test this again with strongly evolving values of the galaxy bias, e.g. $b(z) = \left(\sqrt{1+z}\right)^n$, where $n=-10, 2$, and find this still has little effect on the radialisation of $R^C_\ell(k)$. The growth values for standard concordance cosmology do not seem to have an effect on reaching the radialisation limit.

\subsection{Radialisation Limit}\label{sec:radialisation}

We are interested in quantifying the effect of the radialisation on the BAOs themselves. In Figure \ref{fig:slices}, we consider slices in $\ell$ and $k$ through $R^C_\ell(k)$ in order to investigate how rapidly the limit in Equation \ref{eq:clkpk_obs} is reached. The black (dashed) lines in Figure \ref{fig:slices} correspond to $R^P(k)$ (which are independent of $\ell$ and of the survey selection function) and the red (solid) lines correspond to $R^C_\ell(k)$ for different slices in $\ell$ space and for narrow (LHS) and wide (RHS) selection functions. As the survey selection function is widened, we find that: 
\be \lim_{r_0 \to \infty}R^C_\ell(k)=R^P(k),\ee i.e. $R^C_\ell(k)$ tends towards the oscillations in $R^P(k)$, not only in phase, but also in amplitude. This is noticeable in Figure \ref{fig:clk}: the amplitude of the first peak for example is higher for the deep survey ($>9\%$) than for the shallow survey ($<8\%$). This is another check of Equation \ref{eq:clkpk}.

Equation \ref{eq:radial_limit1} gives us an analytic way to study the limit where we can expect the SFB power spectrum to behave radially. In Figure \ref{fig:radial}, we plot the lines corresponding to the radial limit in $(k,\ell)$ space for different surveys with Gaussian distributions (Equation \ref{eq:phi_gauss}) corresponding to $r_0 = 40, 70, 200, 300, 1100 \left[ h^{-1} {\rm Mpc}\right]$. We have chosen these values as they correspond roughly to the following surveys (for the chosen fiducial survey): IRAS \citep[$z_{\rm mean} \sim 0.015$,][]{IRAS,IRAS2}, 2MRS \citep[2 MASS All-Sky Redshift Survey, $z_{\rm mean}\sim0.03$,][]{2MRS}, SDSS \citep[Sloan Digital Sky Survey, $z_{\rm mean}\sim0.10$,][]{SDSS}, 2dF \citep[2 degree Field, $z_{\rm mean}\sim0.15$][]{2dFdata:2003} and a Stage IV galaxy survey \citep[$z_{\rm mean}\sim0.80$,][]{DETF}.  
Note that, in principle, these surveys would have to be all-sky to obey the radialisation limits plotted in Figure \ref{fig:radial}. 

The area above each diagonal line corresponds the radial limit, i.e. where the radialisation has taken place and Equation \ref{eq:clkpk_obs} holds. We overplot the radial scales corresponding to the first three peaks of the BAOs to show how these will be probed for different survey depths.

By comparing the diagonal lines in Figure \ref{fig:radial} with the BAO `turnover' (i.e. the $\ell$-dependent $k$-scale at which the BAOs switch from being mostly tangential to mostly radial) in Figure \ref{fig:clk}, we notice however that the BAOs seem to become radial before the full SFB power spectrum does, especially for large $\ell$. This surprising effect, further motivates the use of the SFB analysis for BAOs.

\section{Conclusion}\label{sec:conclusion}

In this paper, we have presented a new way to study Baryon Acoustic Oscillations (BAOs) using a Spherical Fourier-Bessel (SFB) decomposition of a wide-field deep galaxy survey. The BAO signal in SFB space can be studied by considering the ratio $R^C_\ell(k)$ of the SFB power spectrum with wiggles $\left(C^{\rm b}_\ell(k)\right)$ to the smooth SFB power spectrum $\left(C_\ell^{\rm no b}(k)\right)$, similarly to what is done in Fourier space. In this decomposition, BAOs can be observed simultaneously in both the radial ($k$) and tangential ($\ell$) modes. 

For a field which is statistically isotropic and homogeneous (SIH), the SFB power spectrum is purely radial, i.e. independent of tangential modes, and reduces to the Cartesian Fourier power spectrum. In the other extreme limit where the field is in a thin shell, we showed that all the SFB information is contained in the tangential modes
and is simply related to the 2D spherical harmonic power spectrum.

Considering the practical case where the field is observed with a radial selection function (thus partially violating the SIH condition), we find that the radial limit can still be reached for selection functions covering a large radial range (see Appendix \ref{app:gauss}), which we refer to as the \emph{radialisation} of the power spectrum. The radialisation will be limited to a region in ($\ell,k$) space corresponding to $k \gg \sqrt{\ell (\ell+1)}/r_0$, where $r_0$ parameterises the radial selection function. We find that  radialisation is a good approximation for these modes even when evolution due to growth and bias are considered, both of which can be considered as a further violation of the SIH condition. To study this limit, we have derived an analytic solution to the window function of a non-evolving, unbiased galaxy field when the radial selection function is a Gaussian in $r$ centered on the observer (Equation \ref{eq:analytic:window}).

We also find the BAOs considered in SFB space radialise as the survey depth is increased, meaning both the phase and amplitude of the BAOs tend towards the Fourier space ratio $R^P(k)$. This means that BAOs for a wide-field shallow survey have smaller amplitude and are spread across the $(\ell,k)$ space, while BAOs for a wide-field deep survey have a larger amplitude and are confined to the radial modes (Figures \ref{fig:clk} and \ref{fig:slices}).  

We study the radial limit analytically (Figure \ref{fig:radial}) and find that it can in principle be observed (for a limited $\ell$-range and for small physical scales) with all-sky surveys with current surveys depths or for future stage-IV surveys.  This suggest that the first BAO peak becomes radial up to $\ell \sim 10$ for an all-sky survey with similar depth to SDSS or 2dF, and up to $\ell \sim 70$ for an all-sky stage IV survey. In practice though, the radialisation might be observable to even higher $\ell$ since we observe that the BAOs radialise more rapidly than the full SFB power spectrum, especially at large $\ell$. Subsequent BAO peaks also become radial for large values of $\ell$ and for shallower surveys, though these may be already be in the non-linear regime. For modes that have become radial, measurements at different $\ell$'s are useful in practice, to reduce measurement errors due to cosmic variance and shot noise.  

We note that we have ignored redshift-space distortions in our analysis though the prescription for these in SFB space is well known, \citep[see for e.g.,][]{Heavens:1995}. These distortions may affect the radialisation as they will introduce mode-mixing, albeit with a distinct signature and are readily described in the SFB basis.  Further $\ell$-mode mixing will occur for incomplete sky coverage, which can be corrected for using the mask geometry.

\begin{acknowledgements}
The authors are grateful to Pirin Erdo\u gdu, Alan Heavens, Ofer Lahav, Fran\c cois Lanusse, Boris Leistedt and Adam Amara for useful discussions about SFB decompositions. The SFB calculations use the discrete spherical Bessel transform (DSBT) presented in \cite{Lanusse:2011} and the authors are grateful to Fran\c cois Lanusse for help implementing this. We extended iCosmo\footnote{\url{www.icosmo.org}} software for our calculations \citep{icosmo1}. This research is in part supported by the Swiss National Science Foundation (SNSF).  
\end{acknowledgements}

\appendix
\section{Radial Limit for a Gaussian Selection Function}\label{app:gauss}
\begin{figure*}[htbp]
   \centering

   \includegraphics[width=9.5cm]{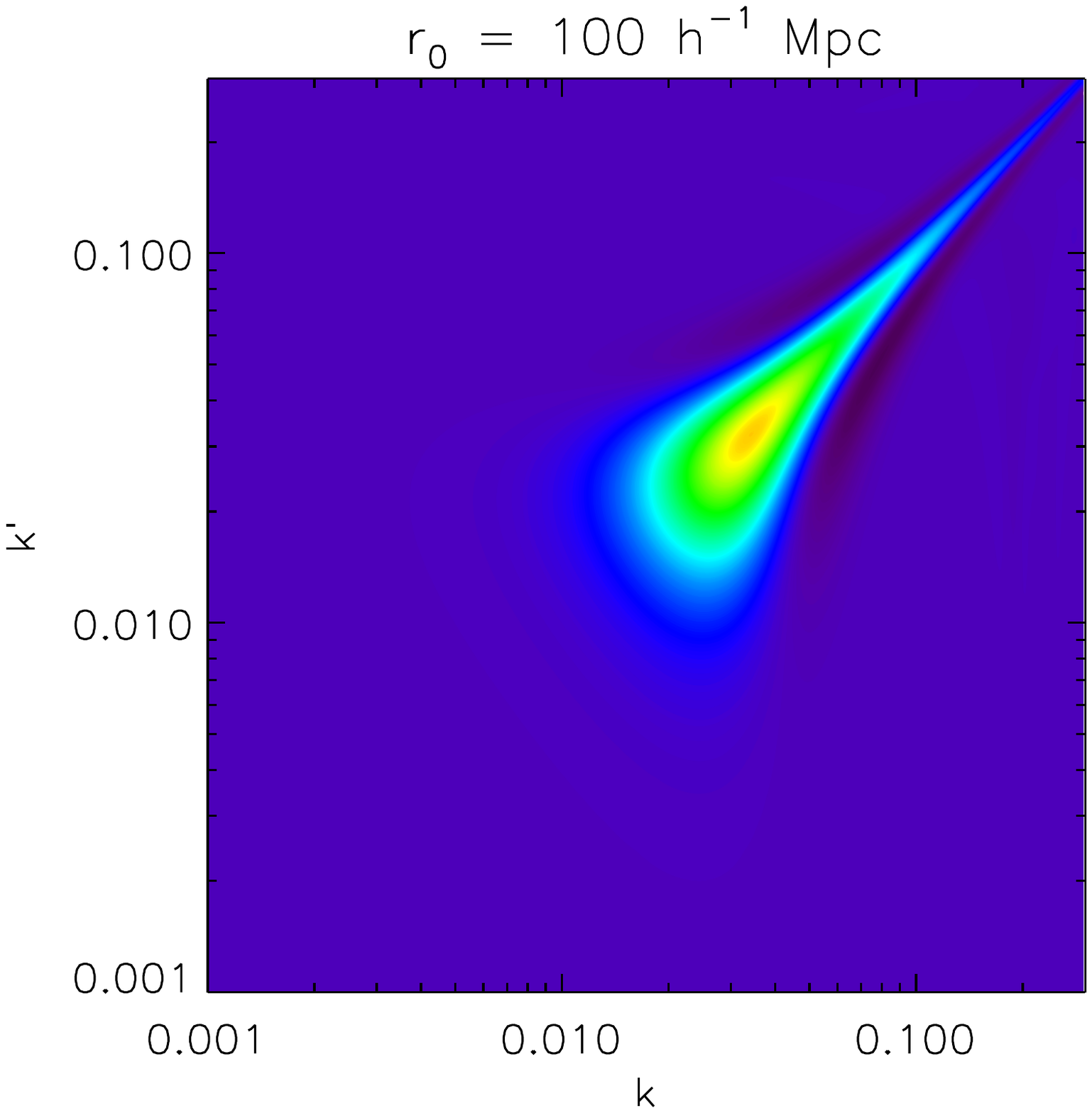}\includegraphics[width=9.5cm]{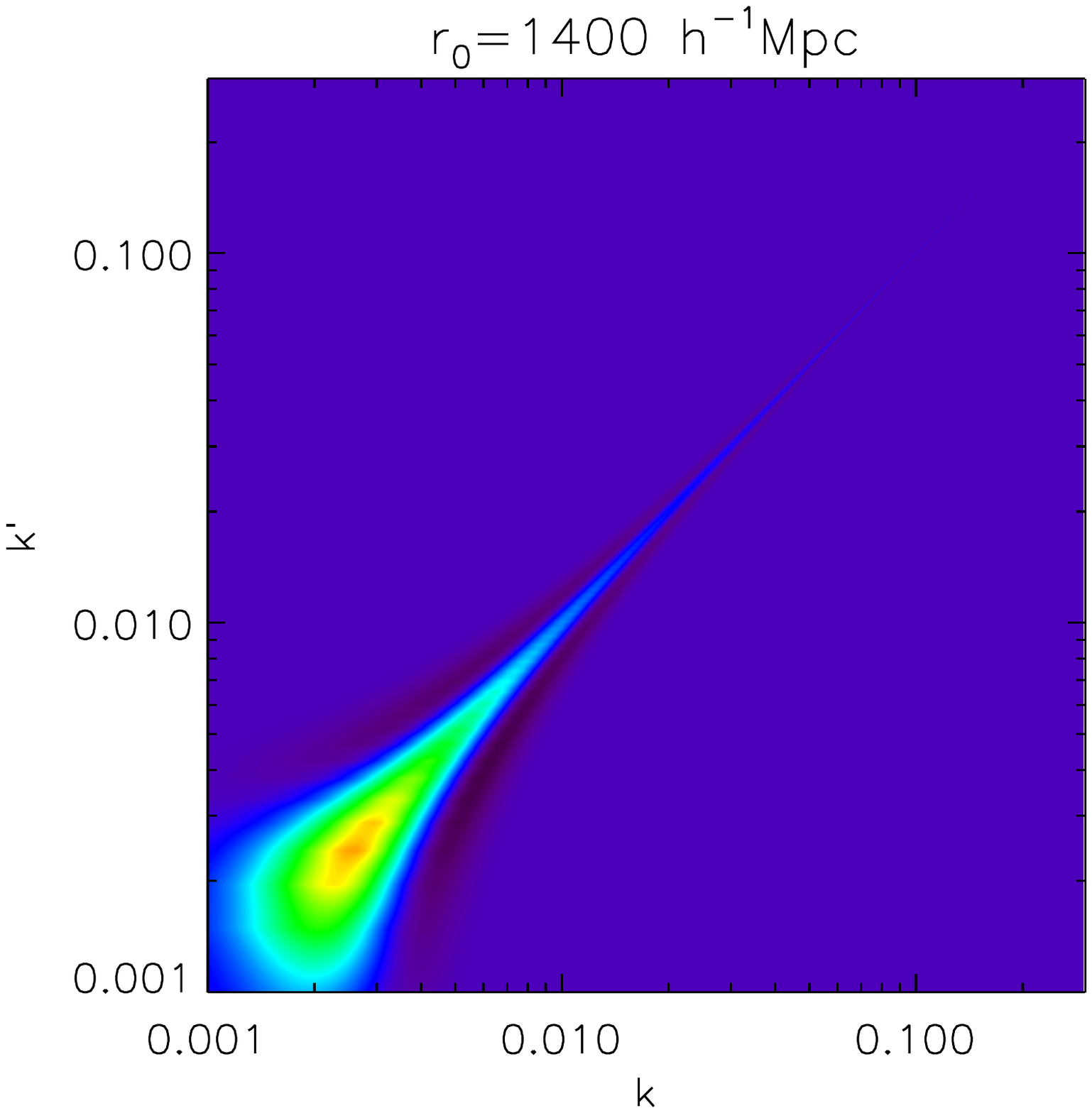}    
   \caption{Window function $W_\ell(k,k')$ for the Gaussian selection function with for $\ell=3$ and $r_0=100 h^{-1}{\rm Mpc}$ (left) and $r_0=1400h^{-1}{\rm Mpc}$ (right).  As the selection function parameter $r_0$ becomes larger (i.e., increasing the redshift coverage from $z_{\rm med}\sim 0.05$ to $z_{\rm med}\sim 0.8$), the window function tends towards  $\frac{\pi}{2k'}\delta(k'-k)$ for scales where $k \gg \sqrt{2\ell(\ell+1)}/r_0$.}
   \label{sec:theory:fig:wg}
\end{figure*}

As shown in \S\ref{subsec:gaussian}, when the galaxy selection is Gaussian, i.e. $\phi(r) =  e^{-(r/r_0)^2}$ , it is the galaxy selection function (Equation~\ref{eq:window}) reduces to 
\bea W_\ell(k,k')&=&\frac{\pi r_0^2}{4}\sqrt{\frac{k}{k'}}\exp\left[-r_0^2\frac{k^2+k'^2}{4}\right]I_{\ell+\frac{1}{2}}\left(
\frac{r_0^2k k'}{2}\right),\nonumber \\~\eea
where $I_\nu(x)$ is the modified Bessel function of the first kind.  For numerical reasons, it is often useful to evaluate the exponentially scaled modified Bessel function of the first kind $\tilde{I_\nu}(x) = \exp(-x)I_\nu(x)$ instead of $I_\nu(x)$, in which case it is useful to re-write the above Equation as: 
\bea W_\ell(k,k')&=&\frac{\pi r_0^2}{4}\sqrt{\frac{k}{k'}}\exp\left[-r_0^2 \left( \frac{k-k'}{2}\right)^2 \right]\tilde{I}_{\ell+\frac{1}{2}}\left(\frac{r_0^2k k'}{2}\right).\nonumber\\~\eea 
Figure~\ref{sec:theory:fig:wg} shows the window function for two values of $r_0$, $\ell=3$ as a function
of $k,k'$. As $r_0$ becomes large, it can be seen that the window function tends towards a delta function
$\frac{1}{k'}\delta(k-k')$.

To study this limit more precisely, we use the asymptotic form for the modified Bessel function
$I_\alpha(x) \simeq e^x/\sqrt{2 \pi x}$ for $x \gg |\alpha^2- 1/4|$, which gives
\begin{equation}
W_{\ell}(k,k') \simeq \frac{\sqrt{\pi}}{4} \frac{r_0}{k'} \exp\left[-r_0^2 \left( \frac{k-k'}{2}\right)^2 \right],
\end{equation}
in the limit where $r_0^2 k k' \gg 2 \ell (\ell+1)$, or approximately,
\begin{equation}
\label{eq:radial_condition}
r_0 k  \gg \sqrt{2 \ell (\ell+1)}.
\end{equation}
Using the definition of a Dirac delta function, namely $\int dx~h(x) \delta(x-x_0) = h(x_0)$ for an arbitrary
function $h(x)$, the window function becomes, in the limit $r_0 \rightarrow \infty$,
\begin{equation}
W_{\ell}(k,k')\simeq \frac{\pi}{2 k'} \delta(k-k'),\label{a5}
\end{equation}
in agreement with Equation~\ref{eq:w_radial} for the radial case discussed in \S\ref{subsec:radial}. Thus, if the condition of Equation~\ref{eq:radial_condition} holds, the power spectrum $C^{\rm obs}_{\ell}(k,k')$ becomes radial
(i.e independent of $\ell$) and equal to $C_\ell(k) \delta(k-k')=P(k)\delta(k-k')$ as in Equation~\ref{eq:clkpk_obs}.

\bibliographystyle{aa}
\bibliography{../../references}

\begin{thebibliography}{32}
\expandafter\ifx\csname natexlab\endcsname\relax\def\natexlab#1{#1}\fi

\bibitem[{{Adelman-McCarthy} {et~al.}(2008){Adelman-McCarthy}, {Ag{\"u}eros},
  {Allam}, {Allende Prieto}, {Anderson}, {Anderson}, {Annis}, {Bahcall},
  {Bailer-Jones}, {Baldry}, {Barentine}, {Bassett}, {Becker}, {Beers}, {Bell},
  {Berlind}, {Bernardi}, {Blanton}, {Bochanski}, {Boroski}, {Brinchmann},
  {Brinkmann}, {Brunner}, {Budav{\'a}ri}, {Carliles}, {Carr}, {Castander},
  {Cinabro}, {Cool}, {Covey}, {Csabai}, {Cunha}, {Davenport}, {Dilday}, {Doi},
  {Eisenstein}, {Evans}, {Fan}, {Finkbeiner}, {Friedman}, {Frieman},
  {Fukugita}, {G{\"a}nsicke}, {Gates}, {Gillespie}, {Glazebrook}, {Gray},
  {Grebel}, {Gunn}, {Gurbani}, {Hall}, {Harding}, {Harvanek}, {Hawley},
  {Hayes}, {Heckman}, {Hendry}, {Hindsley}, {Hirata}, {Hogan}, {Hogg}, {Hyde},
  {Ichikawa}, {Ivezi{\'c}}, {Jester}, {Johnson}, {Jorgensen}, {Juri{\'c}},
  {Kent}, {Kessler}, {Kleinman}, {Knapp}, {Kron}, {Krzesinski}, {Kuropatkin},
  {Lamb}, {Lampeitl}, {Lebedeva}, {Lee}, {Leger}, {L{\'e}pine}, {Lima}, {Lin},
  {Long}, {Loomis}, {Loveday}, {Lupton}, {Malanushenko}, {Malanushenko},
  {Mandelbaum}, {Margon}, {Marriner}, {Mart{\'{\i}}nez-Delgado}, {Matsubara},
  {McGehee}, {McKay}, {Meiksin}, {Morrison}, {Munn}, {Nakajima}, {Neilsen},
  {Newberg}, {Nichol}, {Nicinski}, {Nieto-Santisteban}, {Nitta}, {Okamura},
  {Owen}, {Oyaizu}, {Padmanabhan}, {Pan}, {Park}, {Peoples}, {Pier}, {Pope},
  {Purger}, {Raddick}, {Re Fiorentin}, {Richards}, {Richmond}, {Riess}, {Rix},
  {Rockosi}, {Sako}, {Schlegel}, {Schneider}, {Schreiber}, {Schwope}, {Seljak},
  {Sesar}, {Sheldon}, {Shimasaku}, {Sivarani}, {Smith}, {Snedden}, {Steinmetz},
  {Strauss}, {SubbaRao}, {Suto}, {Szalay}, {Szapudi}, {Szkody}, {Tegmark},
  {Thakar}, {Tremonti}, {Tucker}, {Uomoto}, {Vanden Berk}, {Vandenberg},
  {Vidrih}, {Vogeley}, {Voges}, {Vogt}, {Wadadekar}, {Weinberg}, {West},
  {White}, {Wilhite}, {Yanny}, {Yocum}, {York}, {Zehavi}, \& {Zucker}}]{SDSS}
{Adelman-McCarthy}, J.~K., {Ag{\"u}eros}, M.~A., {Allam}, S.~S., {et~al.} 2008,
  \apjs, 175, 297

\bibitem[{{Albrecht} {et~al.}(2006){Albrecht}, {Bernstein}, {Cahn}, {Freedman},
  {Hewitt}, {Hu}, {Huth}, {Kamionkowski}, {Kolb}, {Knox}, {Mather}, {Staggs},
  \& {Suntzeff}}]{DETF}
{Albrecht}, A., {Bernstein}, G., {Cahn}, R., {et~al.} 2006, ArXiv Astrophysics
  e-prints

\bibitem[{{Cabr{\'e}} \& {Gazta{\~n}aga}(2011)}]{Cabre:Gaz}
{Cabr{\'e}}, A. \& {Gazta{\~n}aga}, E. 2011, \mnras, 412, L98

\bibitem[{{Castro} {et~al.}(2005){Castro}, {Heavens}, \& {Kitching}}]{CHK:2005}
{Castro}, P.~G., {Heavens}, A.~F., \& {Kitching}, T.~D. 2005, \prd, 72, 023516

\bibitem[{{Cole} {et~al.}(2005){Cole}, {Percival}, {Peacock}, {Norberg},
  {Baugh}, {Frenk}, {Baldry}, {Bland-Hawthorn}, {Bridges}, {Cannon}, {Colless},
  {Collins}, {Couch}, {Cross}, {Dalton}, {Eke}, {De Propris}, {Driver},
  {Efstathiou}, {Ellis}, {Glazebrook}, {Jackson}, {Jenkins}, {Lahav}, {Lewis},
  {Lumsden}, {Maddox}, {Madgwick}, {Peterson}, {Sutherland}, \&
  {Taylor}}]{Cole:2005}
{Cole}, S., {Percival}, W.~J., {Peacock}, J.~A., {et~al.} 2005, MNRAS, 362, 505

\bibitem[{{Colless} {et~al.}(2003){Colless}, {Peterson}, {Jackson}, {Peacock},
  {Cole}, {Norberg}, {Baldry}, {Baugh}, {Bland-Hawthorn}, {Bridges}, {Cannon},
  {Collins}, {Couch}, {Cross}, {Dalton}, {De Propris}, {Driver}, {Efstathiou},
  {Ellis}, {Frenk}, {Glazebrook}, {Lahav}, {Lewis}, {Lumsden}, {Maddox},
  {Madgwick}, {Sutherland}, \& {Taylor}}]{2dFdata:2003}
{Colless}, M., {Peterson}, B.~A., {Jackson}, C., {et~al.} 2003, ArXiv
  Astrophysics e-prints

\bibitem[{{Dolney} {et~al.}(2006){Dolney}, {Jain}, \& {Takada}}]{Dolney:2006}
{Dolney}, D., {Jain}, B., \& {Takada}, M. 2006, MNRAS, 366, 884

\bibitem[{{Eisenstein} {et~al.}(2005){Eisenstein}, {Zehavi}, {Hogg},
  {Scoccimarro}, \& et~al}]{Eisenstein:2005}
{Eisenstein}, D.~J., {Zehavi}, I., {Hogg}, D.~W., {Scoccimarro}, R., \& et~al.
  2005, APJ, 633, 560

\bibitem[{{Erdo{\u g}du (a)} {et~al.}(2006)}]{Erdogdu:2005wi}
{Erdo{\u g}du (a)}, P. {et~al.} 2006, Mon. Not. Roy. Astron. Soc., 368, 1515

\bibitem[{{Erdo{\u g}du (b)} {et~al.}(2006){Erdo{\u g}du (b)}, {Lahav},
  {Huchra}, {Colless}, {Cutri}, {Falco}, {George}, {Jarrett}, {Jones}, {Macri},
  {Mader}, {Martimbeau}, {Pahre}, {Parker}, {Rassat}, \&
  {Saunders}}]{Erdogdu:2006dv}
{Erdo{\u g}du (b)}, P., {Lahav}, O., {Huchra}, J.~P., {et~al.} 2006, MNRAS,
  373, 45

\bibitem[{{Fisher} {et~al.}(1995{\natexlab{a}}){Fisher}, {Huchra}, {Strauss},
  {Davis}, {Yahil}, \& {Schlegel}}]{IRAS}
{Fisher}, K.~B., {Huchra}, J.~P., {Strauss}, M.~A., {et~al.}
  1995{\natexlab{a}}, \apjs, 100, 69

\bibitem[{{Fisher} {et~al.}(1995{\natexlab{b}}){Fisher}, {Lahav}, {Hoffman},
  {Lynden-Bell}, \& {Zaroubi}}]{Fisher:1995}
{Fisher}, K.~B., {Lahav}, O., {Hoffman}, Y., {Lynden-Bell}, D., \& {Zaroubi},
  S. 1995{\natexlab{b}}, MNRAS, 272, 885

\bibitem[{{Heavens}(2003)}]{Heavens:2003}
{Heavens}, A. 2003, \mnras, 343, 1327

\bibitem[{{Heavens} \& {Taylor}(1995)}]{Heavens:1995}
{Heavens}, A.~F. \& {Taylor}, A.~N. 1995, MNRAS, 275, 483

\bibitem[{{Huchra} {et~al.}(2011){Huchra}, {Macri}, {Masters}, {Jarrett},
  {Berlind}, {Calkins}, {Crook}, {Cutri}, {Erdogdu}, {Falco}, {George},
  {Hutcheson}, {Lahav}, {Mader}, {Mink}, {Martimbeau}, {Schneider},
  {Skrutskie}, {Tokarz}, \& {Westover}}]{2MRS}
{Huchra}, J.~P., {Macri}, L.~M., {Masters}, K.~L., {et~al.} 2011, ArXiv
  e-prints

\bibitem[{{Kitching} {et~al.}(2011){Kitching}, {Heavens}, \&
  {Miller}}]{KHM:2010}
{Kitching}, T.~D., {Heavens}, A.~F., \& {Miller}, L. 2011, \mnras, 413, 2923

\bibitem[{{Kitching} {et~al.}(2008){Kitching}, {Taylor}, \&
  {Heavens}}]{Kitching:2008}
{Kitching}, T.~D., {Taylor}, A.~N., \& {Heavens}, A.~F. 2008, \mnras, 389, 173

\bibitem[{{Lanusse} {et~al.}(2011){Lanusse}, {Rassat}, \&
  {Starck}}]{Lanusse:2011}
{Lanusse}, F., {Rassat}, A., \& {Starck}, J.-L. 2011, ArXiv e-prints

\bibitem[{{Leistedt} {et~al.}(2011){Leistedt}, {Rassat}, {Refregier}, \&
  {Starck}}]{Leistedt:2011}
{Leistedt}, B., {Rassat}, A., {Refregier}, A., \& {Starck}, J.-L. 2011, ArXiv
  e-prints

\bibitem[{{Peacock} {et~al.}(2006){Peacock}, {Schneider}, {Efstathiou},
  {Ellis}, {Leibundgut}, {Lilly}, \& {Mellier}}]{WGFC}
{Peacock}, J.~A., {Schneider}, P., {Efstathiou}, G., {et~al.} 2006, {ESA-ESO
  Working Group on ''Fundamental Cosmology''}, Tech. rep.

\bibitem[{{Peebles} \& {Yu}(1970)}]{Peebles:1970Y}
{Peebles}, P.~J.~E. \& {Yu}, J.~T. 1970, Apj, 162, 815

\bibitem[{{Percival} {et~al.}(2004){Percival}, {Burkey}, {Heavens}, {Taylor},
  {Cole}, {Peacock}, {Baugh}, {Bland-Hawthorn}, {Bridges}, {Cannon}, {Colless},
  {Collins}, {Couch}, {Dalton}, {De Propris}, {Driver}, {Efstathiou}, {Ellis},
  {Frenk}, {Glazebrook}, {Jackson}, {Lahav}, {Lewis}, {Lumsden}, {Maddox},
  {Norberg}, {Peterson}, {Sutherland}, \& {Taylor}}]{Percival:2004}
{Percival}, W.~J., {Burkey}, D., {Heavens}, A., {et~al.} 2004, \mnras, 353,
  1201

\bibitem[{{Percival} {et~al.}(2007 b){Percival}, {Cole}, {Eisenstein},
  {Nichol}, {Peacock}, {Pope}, \& {Szalay}}]{Percival:2007}
{Percival}, W.~J., {Cole}, S., {Eisenstein}, D.~J., {et~al.} 2007 b, MNRAS,
  381, 1053

\bibitem[{{Rassat} {et~al.}(2008){Rassat}, {Amara}, {Amendola}, {Castander},
  {Kitching}, {Kunz}, {Refregier}, {Wang}, \& {Weller}}]{Rassat:2008bao}
{Rassat}, A., {Amara}, A., {Amendola}, L., {et~al.} 2008, ArXiv e-prints

\bibitem[{{Refregier} {et~al.}(2011){Refregier}, {Amara}, {Kitching}, \&
  {Rassat}}]{icosmo1}
{Refregier}, A., {Amara}, A., {Kitching}, T.~D., \& {Rassat}, A. 2011, \aap,
  528, A33

\bibitem[{{Seo} \& {Eisenstein}(2003)}]{Seo:2003Eis}
{Seo}, H.-J. \& {Eisenstein}, D.~J. 2003, APJ, 598, 720

\bibitem[{{Seo} \& {Eisenstein}(2007)}]{Seo:2007Eis}
{Seo}, H.-J. \& {Eisenstein}, D.~J. 2007, APJ, 665, 14

\bibitem[{{Shapiro} {et~al.}(2011){Shapiro}, {Crittenden}, \&
  {Percival}}]{ISW3D}
{Shapiro}, C., {Crittenden}, R.~G., \& {Percival}, W.~J. 2011, ArXiv e-prints

\bibitem[{{Slosar} {et~al.}(2009){Slosar}, {Ho}, {White}, \&
  {Louis}}]{baoconfig2}
{Slosar}, A., {Ho}, S., {White}, M., \& {Louis}, T. 2009, \jcap, 10, 19

\bibitem[{{Strauss} {et~al.}(1992){Strauss}, {Huchra}, {Davis}, {Yahil},
  {Fisher}, \& {Tonry}}]{IRAS2}
{Strauss}, M.~A., {Huchra}, J.~P., {Davis}, M., {et~al.} 1992, \apjs, 83, 29

\bibitem[{{Sunyaev} \& {Zeldovich}(1970)}]{Sunyaev:1970Z}
{Sunyaev}, R.~A. \& {Zeldovich}, Y.~B. 1970, apss, 7, 3

\bibitem[{{Xu} {et~al.}(2010){Xu}, {White}, {Padmanabhan}, {Eisenstein},
  {Eckel}, {Mehta}, {Metchnik}, {Pinto}, \& {Seo}}]{baoconfig1}
{Xu}, X., {White}, M., {Padmanabhan}, N., {et~al.} 2010, \apj, 718, 1224

\end{thebibliography}

\end{document}